\pdfoutput=1

\documentclass[10pt,conference]{IEEEtran}

\usepackage[utf8]{inputenc}
\usepackage[T1]{fontenc}

\usepackage{amsmath,amssymb,amsfonts,amsthm,mathtools}
\usepackage{booktabs}
\usepackage{csquotes}
\usepackage[noend]{algorithm, algpseudocode}
\floatname{algorithm}{Algorithm}

\usepackage{graphicx}
\usepackage{textcomp}
\usepackage{flushend}
\usepackage[x11names,svgnames,usenames,dvipsnames,table,rgb]{xcolor}
\usepackage[textsize=tiny,loadshadowlibrary,shadow]{todonotes}

\usepackage{braket}
\usepackage{arydshln}
\usepackage{multirow}
\usepackage{float}
\usepackage{color}
\usepackage{soul}
\usepackage{siunitx}
\usepackage{tabularx}
\usepackage{caption}
\usepackage{subcaption}

\usepackage{tikz}
\usetikzlibrary{arrows}
\usetikzlibrary{positioning}  
\usetikzlibrary{shadows}
\usetikzlibrary{arrows,backgrounds,calc,chains,
	matrix,positioning,shapes,shapes.geometric,
	shapes.arrows, decorations.pathmorphing, decorations.pathreplacing}

\tikzstyle{vecArrow} = [thick, decoration={markings,mark=at position
	1 with {\arrow[semithick]{open triangle 60}}}]
 
\usetikzlibrary{positioning,calc,fit,backgrounds,matrix,shapes,arrows.meta,graphs,quotes,automata,shapes.geometric,chains}

\tikzset{%
	font={\footnotesize},
	vertex/.style={draw,circle,inner sep=0pt,minimum width=0.5cm,minimum height=0.5cm,font=\small, scale=0.9},
	terminal/.style={draw,fill=white,rectangle,inner sep=2pt,font=\footnotesize,very thick},
	define color/.code={\definecolor{hsb#1}{Hsb}{#1, 1, 0.75}},
	medge/.style n args={3}{
			line width={#1pt},
			define color={#2},
			draw=hsb#2,
			out=#3,
			in=90
		},
	edge/.style 2 args={
			line width={#1pt},
			define color={#2},
			draw=hsb#2
		},
	edge0/.style 2 args={
			line width={#1pt},
			define color={#2},
			draw=hsb#2,
			out=-130,
			in=90
		},
	edge1/.style 2 args={
			line width={#1pt},
			define color={#2},
			draw=hsb#2,
			out=-50,
			in=90
		},
	zerostub/.style={
			inner sep=0,
			minimum size=3pt,
			circle,
			fill=black
		},
	cross/.style={cross out, draw=black,
			minimum size=2*(#1-\pgflinewidth),
			inner sep=0pt, outer sep=0pt,rotate=45},
	cross/.default={3pt},
	edgeOne/.style={color=RedEdge,ultra thick},
	edgeOneState/.style={color=RedEdge, thick},
	edgeMOne/.style={color=BlueEdge,ultra thick},
	edgeSqrt/.style={color=RedEdge, thick},
	edgeMSqrt/.style={color=BlueEdge, thick},
	edgeFrac/.style={color=RedEdge, thin},
	edgeOver/.style={dotted, color=blue, ultra thick},
	qubit/.style={draw,circle,inner sep=0pt,minimum width=0.35cm,minimum height=0.35cm,font=\footnotesize, thin}
}
\definecolor{RedEdge}{RGB}{191,40,40}
\definecolor{BlueEdge}{RGB}{40,191,191}
\definecolor{Blue01}{rgb}{0.26,0.39,0.85}
\definecolor{Yellow12}{rgb}{1.0,0.88,0.1}
\definecolor{Gray23}{rgb}{0.66,0.66,0.66}
\definecolor{myyellow}{RGB}{255, 253, 0}
\definecolor{myorange}{RGB}{238, 135, 51}
\definecolor{myblue}{RGB}{47, 112, 137}

\makeatletter
\let\MYcaption\@makecaption
\makeatother
\usepackage[font=footnotesize]{subcaption}
\makeatletter
\let\@makecaption\MYcaption
\makeatother

\usepackage{tikz}
\usepackage[compat=0.6]{yquant}
\useyquantlanguage{groups}

\usepackage[
	backend=biber,
	style=ieee,
	sortcites=true,
	url=true,
	eprint=true,
	giveninits=false,
	minnames=1,
	maxnames=3
	]{biblatex}

\usepackage{hyperref}

\newtheorem{example}{Example}

\def\equationautorefname~#1\null{Eq.~(#1)\null}

\AtEveryBibitem{
	\clearname{editor}%
	\clearfield{series}%
	\clearfield{isbn}%
	\clearfield{issn}%
	\clearfield{day}%
	\clearfield{month}%
	\clearfield{place}%
	\clearlist{location}%
	\clearfield{eventtitle}%
	\ifentrytype{inproceedings}{%
		\clearlist{publisher}
	}{}%
	\ifentrytype{software}{%
	}{%
		\clearfield{url}%
		\clearfield{urlyear}%
	}%
}
\begin{document}

\title{Stripping Quantum Decision Diagrams \\ of their Identity}
	\author{
		\IEEEauthorblockN{%
			Aaron Sander\IEEEauthorrefmark{1}\hspace{1cm}%
            Ioan-Albert Florea\IEEEauthorrefmark{1}\hspace{1cm}%
			Lukas Burgholzer\IEEEauthorrefmark{1}\hspace{1cm}%
			Robert Wille\IEEEauthorrefmark{1}\IEEEauthorrefmark{4}%
		}
		\IEEEauthorblockA{\IEEEauthorrefmark{1}Chair for Design Automation, Technical University of Munich, Munich, Germany}
		\IEEEauthorblockA{\IEEEauthorrefmark{4}Software Competence Center Hagenberg (SCCH) GmbH, Hagenberg, Austria}
		\IEEEauthorblockA{
			\href{mailto:aaron.sander@tum.de}{aaron.sander@tum.de},
			\href{mailto:lukas.burgholzer@tum.de}{lukas.burgholzer@tum.de},
			\href{mailto:robert.wille@tum.de}{robert.wille@tum.de}}%
		\IEEEauthorblockA{\url{https://www.cda.cit.tum.de/research/quantum/}\vspace{-.4cm}}
	}

\maketitle

\begin{abstract}
	Classical representations of quantum states and operations as vectors and matrices are plagued by an exponential growth in memory and runtime requirements for increasing system sizes.
	Based on their use in classical computing, an alternative data structure known as \emph{Decision Diagrams} (DDs) has been proposed, which, in many cases, provides both a more compact representation and more efficient computation.
	In the classical realm, decades of research have been conducted on DDs and numerous variations tailored for specific applications exist.
	However, DDs for quantum computing are just in their infancy and there is still room for tailoring them to this new technology.
	In particular, existing representations of DDs require extending all operations in a quantum circuit to the full system size through extension by nodes representing identity matrices.
	In this work, we make an important step forward for quantum DDs by stripping these identity structures from quantum operations. This significantly reduces the number of nodes required to represent them as well as eases the pressure on key building blocks of their implementation.
	As a result, we obtain a structure that is more natural for quantum computing and significantly speeds up computations---with a runtime improvement of up to $70\times$ compared to the state-of-the-art.
\end{abstract}


\section{Introduction}
Quantum computing is a promising new technology that is step-by-step becoming closer to reality and has the chance to propel our computational abilities forward to solve currently intractable problems.
Similar to classical computing, algorithms on quantum computers can be decomposed into smaller operations known as \emph{gates}.
These gates form a \emph{quantum circuit} analogous to digital circuits in which their combined operation on \emph{quantum bits} (qubits) performs some computation.
However, currently it is still necessary to use classical methods to simulate, verify, and compile these circuits.
Unfortunately, these tasks become increasingly difficult due to the standard representation of quantum states and operations (namely vectors and matrices) grows exponentially relative to the number of qubits in the algorithm.
Storing and manipulating these large vectors and matrices quickly becomes infeasible for classical computers---motivating both the need for a quantum computer to perform these computations as well as the need for further development of sophisticated classical methods to represent and work with states and operations in quantum circuits.
Eventually, the capabilities of the most powerful classical simulators define the boundary of quantum advantage.

In the classical realm, the design automation community has spent decades to successfully develop solutions for tackling excessive memory requirements.
One of these solutions is to use \emph{decision diagrams} to represent information.
Over the last decades, a plethora of different types
tailored for different application scenarios
has emerged~\cite{bryantGraphbasedAlgorithmsBoolean1986,minatoZerosuppressedBDDsSet1993,bryantVerificationArithmeticCircuits2001,vandijkTaggedBDDsCombining2017}.
Inspired by their success in the classical realm, decision diagrams have been adapted to the quantum realm in order to exploit both sparsity as well as redundancy in the underlying structures they represent---leading to significant compression in memory requirements and reduction in the runtime necessary to perform calculations~\cite{wangXQDDbasedVerificationMethod2008,abdollahiAnalysisSynthesisQuantum2006,viamontesHighperformanceQuIDDBasedSimulation2004,tsaiBitslicingHilbertSpace2021,millerQMDDDecisionDiagram2006,niemannQMDDsEfficientQuantum2016,hong_tensor_2022,willeDecisionDiagramsQuantum2023,vinkhuijzen_limdd_2023}.
However, these methods do not fully exploit that the underlying representations originate from a quantum context, which
leaves a huge potential untapped.

In this work, we focus on quantum decision diagrams as defined in~\cite{willeDecisionDiagramsQuantum2023}.
This type of DDs always requires the operands of any DD operation (such as multiplication or addition) to act on the same number of qubits---a reasonable assumption considering that, e.g., plain matrix addition also requires both matrices to have the same dimensions.
This is accomplished by blowing up the DD representations of operations with identity nodes for any qubit that is not acted on.
Since most quantum operations only feature a low number of qubits (typically one or two), this incurs a substantial overhead---not only in the sheer number of nodes but also in the pressure that is being put on key data structures such as compute and unique tables within the respective DD package.
Additionally, these identity structures inherently do not play a role in the circuit as they do not perform any action.

Motivated by this fact, this work proposes a new DD structure that strips away these identities---leading to a significantly more compact representation that, simultaneously, better mimics the quantum gates that it represents.
Experimental evaluations on DD-based statevector and unitary simulation demonstrate that this seemingly simple change has profound implications on the performance of the resulting DD package---resulting in an average speed-up of $7.7\times$ and up to an $70\times$ improvement compared to the state of the art.
The resulting implementation is publicly available at \url{https://github.com/cda-tum/mqt-ddsim} as part of the \emph{Munich Quantum Toolkit} 
(MQT;\cite{wille_mqt_2024}).


The rest of this paper is structured as follows:
\autoref{sec:QuantumComputing} reviews the basics of quantum computing and decision diagrams.
Based on that,~\autoref{sec:Motivation} motivates the proposed idea of an identity-less DD structure---with detailed implementation changes described in~\autoref{sec:Implementation} and implications of this change discussed in~\autoref{sec:implications}.
Afterwards,~\autoref{sec:Evaluation} presents and discusses the obtained experimental results, before~\autoref{sec:Conclusion} concludes the paper.


\section{Background} \label{sec:QuantumComputing}
In order to keep this paper self-contained, this section briefly covers the basics of quantum computing used in the remainder of this work and reviews decision diagrams for representing and manipulating quantum states and operations.


\subsection{Quantum Computing}
While classical computing relies on bits (that can either be $0$ or $1$), quantum computing relies on \emph{quantum bits} (or \emph{qubits}) that can also be $\ket{0}$ or $\ket{1}$, but additionally in an arbitrary \emph{superposition} of both computational basis states.
The state $\ket{\Psi}$ of a single qubit is described as $\alpha_0 \ket{0} + \alpha_1 \ket{1}$, with \mbox{complex-valued} \emph{amplitudes} $\alpha_0$ and $\alpha_{1}$ such that $|\alpha_0|^2 + |\alpha_1|^2 = 1$.
In general, the state $\ket{\Psi}$ of an $n$-qubit system is described by $2^n$ \mbox{complex-valued} amplitudes~$\alpha_{i}$ that describe a linear combination of the computational basis states $\ket{i}$ for $i=0,\dots,2^n-1$.
Here, $\ket{i}$ can be thought of as the (classical) state corresponding to the bitstring of size $n$ that is given by the binary expansion of the integer $i$.
Again, these amplitudes are normalized such that $\sum_i |\alpha_i|^2 = 1$.
A quantum state $\ket{\Psi}$ is typically represented by a \mbox{complex-valued} vector containing its amplitudes that is frequently referred to as the \emph{statevector}.
\begin{example} \label{ex:BellState}
	Consider the following two-qubit quantum state $\ket{\Psi} = \frac{1}{\sqrt{2}} \bigl( \ket{00} + \ket{11} \bigr)$.
	Then, the corresponding statevector is given by
	\begin{equation}
		\ket{\Psi} = \begin{pmatrix} \frac{1}{\sqrt{2}} & 0 & 0 & \frac{1}{\sqrt{2}}\end{pmatrix}^T.
	\end{equation}
	This state is known as a \emph{Bell state} and is the smallest example of an \emph{entangled} quantum state, where the state of the individual qubits cannot be described separately any more.
\end{example}

Similar to quantum states being represented by vectors, \emph{quantum operations} (also called \emph{quantum gates}) are represented by matrices that are unitary, i.e., $U^\dag U = I$ with $U^\dag$ denoting the conjugate transpose of $U$ and $I$ denoting the identity matrix.
While \mbox{$n$-qubit} states require $2^n$ complex values, $n$-qubit operations require $2^n \times 2^n$ entries.

\begin{example}\label{ex:Operations}
	Common examples of single-qubit gates are the Pauli-$X$ and $H$ (Hadamard) gates. The $X$ gate is analogous to a bit-flip, while the $H$ gate is used to generate a superposition from a computational basis state.
	These, along with the identity operation, are defined in matrix form as follows
	\begin{equation}
		X = \begin{pmatrix} 0 & 1 \\
			1 & 0 \end{pmatrix},
		H = \frac{1}{\sqrt{2}}\begin{pmatrix} 1 & 1 \\
			1 & -1 \end{pmatrix}, \textit{ and }
		I = \begin{pmatrix} 1 & 0 \\
			0 & 1 \end{pmatrix}.
	\end{equation}
	One of the most common two-qubit gates is the \emph{\mbox{Controlled-NOT}} (CNOT) gate, which applies an $X$ gate to a \emph{target} qubit conditioned on a \emph{control} qubit being $\ket{1}$.
	This corresponding unitary matrix is given by
	\begin{equation} \label{eq:CNOT}
		CNOT = \begin{pmatrix} 1 & 0 & 0 & 0 \\
			0 & 1 & 0 & 0 \\
			0 & 0 & 0 & 1 \\
			0 & 0 & 1 & 0 \end{pmatrix}
		= \begin{pmatrix} I & 0 \\
			0 & X \end{pmatrix},
	\end{equation}
	which, as shown, is equivalent to smaller $2 \times 2$ blocks corresponding to the identity $I$ and $X$ gate.
\end{example}

In general, a \emph{quantum algorithm} is a unitary transformation and, hence, can itself be represented as a unitary matrix $U$ that encodes the full functionality of the algorithm.
The application of a quantum algorithm to a certain initial state is then conceptually equivalent to the matrix-vector multiplication of the algorithm's unitary $U$ and the quantum state's statevector~$\ket{\Psi}$, i.e., $\ket{\Psi'} = U \ket{\Psi}$ with $\alpha'_i = \sum_{j} u_{ij} \alpha_j \ket{j}$.

%
Since the size of these unitaries grows exponentially with the system size, it is hardly feasible and practicable to represent quantum algorithms in this form.
Even more so, since actual quantum computers only offer a limited (yet universal) set of natively available gates that is typically limited to single- and two-qubit operations.
As a consequence, quantum algorithms are predominantly described as sequences of smaller quantum gates that form a \emph{quantum circuit}.
A quantum algorithm and its circuit representation can be thought of as a direct analogue to high-level classical computations (such as addition) and the logic circuits representing them (such as an adder circuit).

\begin{example} \label{eq:BellStateDecomposition}
	Consider the following quantum circuit $G$
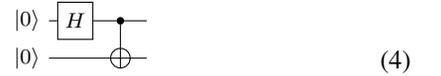
	\begin{equation}\centering
			\begin{tikzpicture}
				\begin{yquant}
						qubit {$\ket{0}$} q[2];
						box {$H$} (q[0]);
						cnot q[1] | q[0];
				\end{yquant}
		\end{tikzpicture}
	\end{equation}
	that acts on two qubits and consists of two gates---a Hadamard gate applied to the top qubit and a CNOT gate controlled by the top qubit and targeted at the bottom qubit.
	Then, the functionality $U$ of $G$ is described by $U = (CNOT)(H \otimes I)$, where $\otimes$ corresponds to the \emph{Kronecker product}, used here to expand the $H$ operation to the full system size with the identity matrix $I$ so that the matrix-matrix multiplication can be applied.
	This results in the unitary
	\begin{equation}
		U = \frac{1}{\sqrt{2}}\begin{pmatrix} 1 & 0 & 1 & 0 \\
							0 & 1 & 0 & 1 \\
							0 & 1 & 0 & -1 \\
							1 & 0 & -1 & 0 \end{pmatrix}.
	\end{equation}
	Applying this unitary to the all-zero initial state $\ket{00}$, i.e., computing
	\begin{equation}
		U \ket{00} = (CNOT)(H \otimes I) \ket{00} = \frac{1}{\sqrt{2}} (\ket{00} + \ket{11})
	\end{equation}
	yields the Bell state from~\autoref{ex:BellState}.
\end{example}

While vectors and matrices are perfectly suitable for representing small-scale quantum systems on classical computers, the inherent exponential complexity quickly becomes prohibitive for larger system sizes.
This motivates the need for alternative methods to efficiently represent and manipulate quantum states and operations on classical computers, thus continuously pushing the boundary of what can currently be simulated and understood without an actual quantum computer.

\subsection{Decision Diagrams}
\emph{Decision Diagrams} (DDs) have been proposed as one such alternative data structure~\cite{wangXQDDbasedVerificationMethod2008,abdollahiAnalysisSynthesisQuantum2006,viamontesHighperformanceQuIDDBasedSimulation2004,tsaiBitslicingHilbertSpace2021,millerQMDDDecisionDiagram2006,niemannQMDDsEfficientQuantum2016,hongTensorNetworkBased2020,willeDecisionDiagramsQuantum2023,vinkhuijzen_limdd_2023}.
Inspired by their classical counterparts, commonly used to represent and manipulate Boolean functions in classical circuit design~\cite{bryantGraphbasedAlgorithmsBoolean1986,minatoZerosuppressedBDDsSet1993,bryantVerificationArithmeticCircuits2001,vandijkTaggedBDDsCombining2017},
the use of the underlying principles has recently been introduced as a tool for classical simulation, verification, and compilation of quantum circuits \cite{zulehnerAdvancedSimulationQuantum2019,burgholzerAdvancedEquivalenceChecking2021,wangXQDDbasedVerificationMethod2008,abdollahiAnalysisSynthesisQuantum2006,viamontesHighperformanceQuIDDBasedSimulation2004,tsaiBitslicingHilbertSpace2021,burgholzerHybridSchrodingerFeynmanSimulation2021,smithQuantumLogicSynthesis2019,willeBasisDesignTools2022,grurlNoiseawareQuantumCircuit2023,sander_towards_2023}.
In the following, we explicitly focus on DDs as described in~\cite{willeDecisionDiagramsQuantum2023} as the basis of this work\footnote{Due to page limitations, the following descriptions had to be kept rather brief. We refer the interested reader to~\cite{willeDecisionDiagramsQuantum2023} and the references therein for an in-depth introduction to decision diagrams.}.
Thereby, the main principle is the recursive subdivision of the underlying representations into subcomponents corresponding to individual qubits and explicitly exploiting sparsity and redundancy throughout this subdivision in conjunction with suitable normalization criteria.

For quantum states, this amounts to recursively halving the corresponding statevector until only scalar numbers remain.
At each division, a node with two successors is introduced---the left successor representing the top half of the (sub)vector and the right successor representing the bottom half of the (sub)vector.
In this splitting, the left successor always leads to an amplitude where the qubit corresponding to the current level in the DD is $\ket{0}$, whereas the right successor leads to amplitudes where that qubits is $\ket{1}$.
Whenever a node is solely composed of zero-entries, it is removed and replaced by a \emph{so-called} \emph{zero-stub} indicating that anything along the respective path will lead to $0$.
DDs are a canonic data structure, so whenever two nodes have an identical structure, only one of them is ever actually represented and shared within a DD---reducing the overall resources required to represent the state. 

\begin{figure}
	\centering
	\begin{subfigure}[b]{0.49\linewidth}
		\centering
		\begin{tikzpicture}[node distance=0.5 and 1.0]
			\node[vertex] (q2) {$q_1$};
			\node[vertex,below left=0.5 and 0.25 of q2] (q1a) {$q_0$};
			\node[vertex,below right=0.5 and 0.25 of q2] (q1b) {$q_0$};
			\node[terminal, below= of q1a] (ta) {};
			\node[terminal, below= of q1b] (tb) {};

		\draw[edge={1}{0}] ($(q2)+(0,0.5cm)$) -- node[midway, left] {$\frac{1}{\sqrt{2}}$} (q2);
		
		\draw[edge0={1}{0}] (q2) to (q1a);
			\draw[edge1={1}{0}] (q2) to (q1b);
		
			\draw[edge1={1}{0}] (q1a) to ++(-50:0.35)  node[zerostub] {};
			\draw[edge0={1}{0}] (q1b) to ++(-140:0.35)  node[zerostub] {};
			
			\draw[edge0={1}{0}] (q1a) to (ta);
			\draw[edge1={1}{0}] (q1b) to (tb);
		\end{tikzpicture}
		\caption{Bell State}
		\label{fig:BellState_DD}
	\end{subfigure}
	\begin{subfigure}[b]{0.49\linewidth}
		\centering
		\begin{tikzpicture}[node distance=0.5 and 0.125]
			\node[vertex] (top) {$q_1$};
			\draw[edge={1}{0}] ($(top)+(0,0.5cm)$) -- (top);
			\node[below left=0.5 and 0.25 of top, vertex] (id) {$q_0$};
			\node[below right=0.5 and 0.25 of top, vertex] (x) {$q_0$};
			\node[terminal,below left=of id] (t0) {};
			\node[terminal,below right=of id] (t1) {};
			\node[terminal,below left=of x] (tx0) {};
			\node[terminal,below right=of x] (tx1) {};
			
			\draw[medge={1}{0}{-130}] (top) to (id);
			\draw[medge={1}{0}{-100}] (top) to ++(-100:0.35)  node[zerostub] {};
			\draw[medge={1}{0}{-80}] (top) to ++(-80:0.35)  node[zerostub] {};
			\draw[medge={1}{0}{-50}] (top) to (x);
	
			\draw[medge={1}{0}{-130}] (id) to (t0);
			\draw[medge={1}{0}{-100}] (id) to ++(-100:0.35)  node[zerostub] {};
			\draw[medge={1}{0}{-80}] (id) to ++(-80:0.35)  node[zerostub] {};
			\draw[medge={1}{0}{-50}] (id) to (t1);
	
			\draw[medge={1}{0}{-130}] (x) to ++(-130:0.35)  node[zerostub] {};
			\draw[medge={1}{0}{-100}] (x) to (tx0);
			\draw[medge={1}{0}{-80}] (x) to (tx1);
			\draw[medge={1}{0}{-50}] (x) to ++(-50:0.35)  node[zerostub] {};
			\end{tikzpicture}
		\caption{CNOT}
		\label{fig:CNOT_DD}
	\end{subfigure}
	\caption{DDs for~\autoref{ex:BellState} and~\autoref{eq:CNOT}}
\end{figure}

\begin{example}
	The Bell state described in \autoref{ex:BellState} is represented as the decision diagram~\autoref{fig:BellState_DD}
	Herein, each level corresponds to a qubit in the system.
	Individual amplitudes are obtained by multiplying the edge weights throughout the tree along the path of a given computational state.
	For example, the amplitude of the $\ket{00}$ state is reconstructed starting at the top of the above DD and going left twice, leading to the computation $1\times \frac{1}{\sqrt{2}}\times 1 = \frac{1}{\sqrt{2}}$.
\end{example}

The decomposition of matrices follows a similar scheme in that the underlying unitary matrix is recursively quartered and nodes with four successors are created at each division.
Here, the left-most successor corresponds to the top-left, the second to the top-right, the third to the bottom-left, and the right-most to the bottom-right quadrant.

\begin{example}
	The CNOT gate is represented by the decision diagram in~\autoref{fig:CNOT_DD}
	Again, each level corresponds to an interaction with a given qubit in the system.
	This is equivalent to the block decomposition as seen in~\autoref{eq:CNOT}.
\end{example}

The unique selling point of DDs is that, instead of scaling with the number of entries in the underlying vectors or matrices, operations on decision diagrams (such as addition and multiplication) scale with the number of nodes in the respective DDs---a direct consequence of their recursive definition.
That is, as long as these representations stay compact, DDs not only allow to compactly represent, but also to efficiently manipulate components relevant for classically conducting quantum computations.

\section{Motivation and General Idea} \label{sec:Motivation}

The above description might make it seem that DDs for quantum computing are a rather mature data structure where everything is solved.
However, compared to the decades of research on variations of classical DDs for dedicated classes of problems, theoretical bounds on their growth, as well as highly engineered software implementations, DDs for quantum computing are still in their infancy with many unanswered questions and lots of potential to improve the underlying concepts.
Especially, since these methods do not fully exploit that the underlying representations originate from a quantum context.
An example illustrates the untapped potential.
\makeatletter
\DeclareRobustCommand\rvdots{%
\vbox{%
\baselineskip4\p@\lineskiplimit\z@%
\kern-\p@%
\hbox{.}\hbox{.}\hbox{.}%
}%
}

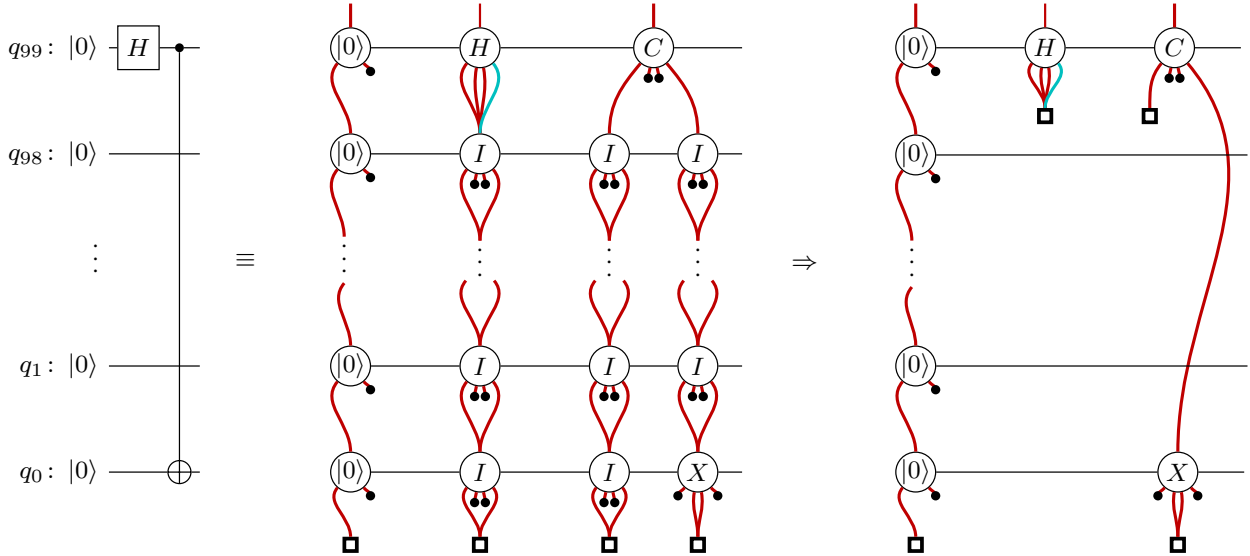
\begin{figure*}
	\centering
	\resizebox{0.95\linewidth}{!}{
	\begin{tikzpicture}
		\begin{yquant}[operator/minimum width=0pt, register/minimum height=3mm,
register/minimum depth=8mm]
					[name=q4]
					qubit {$q_{99}\colon \ket{0}$} a[1];
					[name=q3]
					qubit {$q_{98}\colon \ket{0}$} a[+1];
					[name=q2]
					qubit {$\rvdots$} b[1]; discard b;
					[name=q1]
					qubit {$q_1\colon \ket{0}$} q[1];
					[name=q0]
					qubit {$q_0\colon \ket{0}$} q[+1];
					box {$H$} (a[0]);
					cnot q[1] | a[0];
		\end{yquant}
		\node[right=2.5cm of q4, vertex] (s4) {$\ket{0}$};
		\draw[edge={1}{0}] ($(s4)+(0,0.5cm)$) -- (s4);
		\node[right=2.5cm of q3, vertex] (s3) {$\ket{0}$};
		\draw[edge0={1}{0}] (s4) to (s3);
		\draw[edge1={1}{0}] (s4) to ++(-50:0.35) node[zerostub] {};	
		\node[right=2.5cm of q2] (s2) {$\rvdots$};
		\draw[edge0={1}{0}] (s3) to (s2);
		\draw[edge1={1}{0}] (s3) to ++(-50:0.35) node[zerostub] {};
		\node[right=2.5cm of q1, vertex] (s1) {$\ket{0}$};
		\draw[edge0={1}{0}] (s2.south) to (s1);
		\node[right=2.5cm of q0, vertex] (s0) {$\ket{0}$};
		\draw[edge0={1}{0}] (s1) to (s0);
		\draw[edge1={1}{0}] (s1) to ++(-50:0.35) node[zerostub] {};	
		\node[terminal, below=0.5of s0] (st) {};
		\draw[edge0={1}{0}] (s0) to (st);
		\draw[edge1={1}{0}] (s0) to ++(-50:0.35) node[zerostub] {};	
		\node[] at ($(q2)!0.6!(s2)$) {$\equiv$};
		\node[right=1cm of s4, vertex] (h4) {$H$};
		\draw[edge={0.707}{0}] ($(h4)+(0,0.5cm)$) -- (h4);
		\node[right=1cm of s3, vertex] (h3) {$I$};
		\draw[medge={1}{0}{-130}] (h4) to (h3);
		\draw[medge={1}{0}{-100}] (h4) to (h3);
		\draw[medge={1}{0}{-80}] (h4) to (h3);
		\draw[medge={1}{180}{-50}] (h4) to (h3);
		\draw[medge={1}{0}{-100}] (h3) to ++(-100:0.35)  node[zerostub] {};
		\draw[medge={1}{0}{-80}] (h3) to ++(-80:0.35)  node[zerostub] {};
		\node[below=0.75 of h3] (h2) {$\rvdots$};
		\draw[medge={1}{0}{-130}] (h3) to (h2);
		\draw[medge={1}{0}{-50}] (h3) to (h2);
		\node[right=1cm of s1, vertex] (h1) {$I$};
		\draw[medge={1}{0}{-130}] (h2) to (h1);
		\draw[medge={1}{0}{-50}] (h2) to (h1);
		\draw[medge={1}{0}{-100}] (h1) to ++(-100:0.35)  node[zerostub] {};
		\draw[medge={1}{0}{-80}] (h1) to ++(-80:0.35)  node[zerostub] {};
		\node[right=1cm of s0, vertex] (h0) {$I$};
		\draw[medge={1}{0}{-130}] (h1) to (h0);
		\draw[medge={1}{0}{-50}] (h1) to (h0);
		\draw[medge={1}{0}{-100}] (h0) to ++(-100:0.35)  node[zerostub] {};
		\draw[medge={1}{0}{-80}] (h0) to ++(-80:0.35)  node[zerostub] {};
		\node[terminal, below=0.5of h0] (ht) {};
		\draw[medge={1}{0}{-130}] (h0) to (ht);
		\draw[medge={1}{0}{-50}] (h0) to (ht);
		\node[right=1.5cm of h4, vertex] (cx4) {$C$};
		\draw[edge={1}{0}] ($(cx4)+(0,0.5cm)$) -- (cx4);
		\draw[medge={1}{0}{-100}] (cx4) to ++(-100:0.35)  node[zerostub] {};
		\draw[medge={1}{0}{-80}] (cx4) to ++(-80:0.35)  node[zerostub] {};
		
		\node[right=1cm of h3, vertex] (cx3a) {$I$};
		\draw[medge={1}{0}{-130}] (cx4) to (cx3a);
		\draw[medge={1}{0}{-100}] (cx3a) to ++(-100:0.35)  node[zerostub] {};
		\draw[medge={1}{0}{-80}] (cx3a) to ++(-80:0.35)  node[zerostub] {};
		\node[right=2cm of h3, vertex] (cx3b) {$I$};
		\draw[medge={1}{0}{-50}] (cx4) to (cx3b);
		\draw[medge={1}{0}{-100}] (cx3b) to ++(-100:0.35)  node[zerostub] {};
		\draw[medge={1}{0}{-80}] (cx3b) to ++(-80:0.35)  node[zerostub] {};
		
		\node[below=0.75 of cx3a] (cx2a) {$\rvdots$};
		\draw[medge={1}{0}{-130}] (cx3a) to (cx2a);
		\draw[medge={1}{0}{-50}] (cx3a) to (cx2a);
		
		\node[below=0.75 of cx3b] (cx2b) {$\rvdots$};
		\draw[medge={1}{0}{-130}] (cx3b) to (cx2b);
		\draw[medge={1}{0}{-50}] (cx3b) to (cx2b);
		
		\node[right=1cm of h1, vertex] (cx1a) {$I$};
		\draw[medge={1}{0}{-100}] (cx1a) to ++(-100:0.35)  node[zerostub] {};
		\draw[medge={1}{0}{-80}] (cx1a) to ++(-80:0.35)  node[zerostub] {};
		\draw[medge={1}{0}{-130}] (cx2a) to (cx1a);
		\draw[medge={1}{0}{-50}] (cx2a) to (cx1a);
		\node[right=2cm of h1, vertex] (cx1b) {$I$};
		\draw[medge={1}{0}{-100}] (cx1b) to ++(-100:0.35)  node[zerostub] {};
		\draw[medge={1}{0}{-80}] (cx1b) to ++(-80:0.35)  node[zerostub] {};
		\draw[medge={1}{0}{-130}] (cx2b) to (cx1b);
		\draw[medge={1}{0}{-50}] (cx2b) to (cx1b);
		
		\node[right=1cm of h0, vertex] (cx0a) {$I$};
		\draw[medge={1}{0}{-100}] (cx0a) to ++(-100:0.35)  node[zerostub] {};
		\draw[medge={1}{0}{-80}] (cx0a) to ++(-80:0.35)  node[zerostub] {};
		\draw[medge={1}{0}{-130}] (cx1a) to (cx0a);
		\draw[medge={1}{0}{-50}] (cx1a) to (cx0a);
		\node[right=2cm of h0, vertex] (cx0b) {$X$};
		\draw[medge={1}{0}{-130}] (cx0b) to ++(-130:0.35)  node[zerostub] {};
		\draw[medge={1}{0}{-50}] (cx0b) to ++(-50:0.35)  node[zerostub] {};
		\draw[medge={1}{0}{-130}] (cx1b) to (cx0b);
		\draw[medge={1}{0}{-50}] (cx1b) to (cx0b);
		
		\node[terminal, below=0.5 of cx0a] (cxta) {};
		\draw[medge={1}{0}{-130}] (cx0a) to (cxta);
		\draw[medge={1}{0}{-50}] (cx0a) to (cxta);
		\node[terminal, below=0.5 of cx0b] (cxtb) {};
		\draw[medge={1}{0}{-100}] (cx0b) to (cxtb);
		\draw[medge={1}{0}{-80}] (cx0b) to (cxtb);
		\node[right=2.5cm of cx4, vertex] (z4) {$\ket{0}$};
		\draw[edge={1}{0}] ($(z4)+(0,0.5cm)$) -- (z4);
		\node[below=0.75 of z4, vertex] (z3) {$\ket{0}$};
		\draw[edge0={1}{0}] (z4) to (z3);
		\draw[edge1={1}{0}] (z4) to ++(-50:0.35) node[zerostub] {};	
		\node[right=2.1cm of cx2b] (z2) {$\rvdots$};
		\draw[edge0={1}{0}] (z3) to (z2);
		\draw[edge1={1}{0}] (z3) to ++(-50:0.35) node[zerostub] {};
		\node[right=2cm of cx1b, vertex] (z1) {$\ket{0}$};
		\draw[edge0={1}{0}] (z2.south) to (z1);
		\node[right=2cm of cx0b, vertex] (z0) {$\ket{0}$};
		\draw[edge0={1}{0}] (z1) to (z0);
		\draw[edge1={1}{0}] (z1) to ++(-50:0.35) node[zerostub] {};	
		\node[terminal, below=0.5of z0] (zt) {};
		\draw[edge0={1}{0}] (z0) to (zt);
		\draw[edge1={1}{0}] (z0) to ++(-50:0.35) node[zerostub] {};
		\node[right=1cm of z4, vertex] (hh4) {$H$};
		\draw[edge={0.707}{0}] ($(hh4)+(0,0.5cm)$) -- (hh4);
		\node[terminal, below=0.45of hh4] (ht) {};
		\draw[medge={1}{0}{-130}] (hh4) to (ht);
		\draw[medge={1}{0}{-100}] (hh4) to (ht);
		\draw[medge={1}{0}{-80}] (hh4) to (ht);
		\draw[medge={1}{180}{-50}] (hh4) to (ht);
		\node[right=1cm of hh4, vertex] (cxx4) {$C$};
		\draw[edge={1}{0}] ($(cxx4)+(0,0.5cm)$) -- (cxx4);
		\draw[medge={1}{0}{-100}] (cxx4) to ++(-100:0.35)  node[zerostub] {};
		\draw[medge={1}{0}{-80}] (cxx4) to ++(-80:0.35)  node[zerostub] {};
		\node[terminal, right=1cm of ht] (cxxt) {};
		\draw[medge={1}{0}{-130}] (cxx4) to (cxxt);
		\node[right=2.5cm of z0, vertex] (targ) {$X$};
		\node[terminal, below=0.5 of targ] (term) {};
		\draw[medge={1}{0}{-100}] (targ) to (term);
		\draw[medge={1}{0}{-80}] (targ) to (term);
		\draw[medge={1}{0}{-130}] (targ) to ++(-130:0.35)  node[zerostub] {};
		\draw[medge={1}{0}{-50}] (targ) to ++(-50:0.35)  node[zerostub] {};
		\draw[medge={1}{0}{-50}] (cxx4) to (targ);
		\node[] at ($(cx2b)!0.5!(z2)$) {$\Rightarrow$};
		
		\draw[] (s0) -- (h0) -- (cx0a) -- (cx0b) -- ($(cx0b)+(0.5, 0)$);
		\draw[] (s1) -- (h1) -- (cx1a) -- (cx1b) -- ($(cx1b)+(0.5, 0)$);
		\draw[] (s3) -- (h3) -- (cx3a) -- (cx3b) -- ($(cx3b)+(0.5, 0)$);
		\draw[] (s4) -- (h4) -- (cx4) -- ($(cx4)+(1.0, 0)$);
		
		\draw[] (z4) -- (hh4) -- (cxx4) -- ($(cxx4)+(0.75, 0)$);
		\draw[] (z0) -- (targ) -- ($(targ)+(0.75, 0)$);
		\draw[] (z3) -- ($(z3)+(3.75, 0)$);
		\draw[] (z1) -- ($(z1)+(3.75, 0)$);
		
	\end{tikzpicture}}
	\caption{Bell State between $0^{\text{th}}$ and $99^{\text{th}}$ qubit as described in~\autoref{ex:LargeBellState} represented as a circuit, the current DD structure, and the new identity-less structure}
	\label{fig:BellState_NewDD}
\end{figure*}

\begin{example} \label{ex:LargeBellState}
	Say that we have a $100$-qubit system and want to generate a Bell state between the first and the last qubit such that we generate the state
	\begin{equation}
		\ket{\Psi} = \frac{1}{\sqrt{2}} \bigl( \ket{0 \underbrace{\dots 0 \dots}_{\text{98 times}} 0} + \ket{1 \underbrace{\dots 0 \dots}_{\text{98 times}} 1} \bigr)
	\end{equation}
	The circuit used to generate this state as well as the DD representing every entity in it are shown in~\autoref{fig:BellState_NewDD}.

	Observe how each gate DD contains identity nodes at levels which are not affected by the operation (drawn as wires).
	This is a direct consequence of the extension to the full system size previously observed in \autoref{eq:BellStateDecomposition} that is used to make the dimensions of the respective quantities fit.
	As a consequence, the DD for the single-qubit Hadamard gate consists of $100$ nodes while the DD for the CNOT gate is blown up to a total of $199$ nodes (one at the control, 2 identities at each intermediate level, plus an identity and an X gate at the target level).
\end{example}

As the above example has shown, the extension to the full system size introduces a severe overhead for representing quantities that per-definition do not affect the circuit functionality at all.
For that reason, many sophisticated implementations of other techniques for statevector simulation directly manipulate the amplitudes of the state vector that are affected by an operation and never construct the full operation matrix (which, in many cases, could not even be feasibly represented due to its exponential size)~\cite{deraedtMassivelyParallelQuantum2019,hanerPetabyteSimulation45Qubit2017,guerreschiIntelQuantumSimulator2020}.
To further advance the state of the art in decision diagrams for quantum computing, it is necessary to strip decision diagrams representing quantum operations of their identity nodes.
This makes them both a more natural representation of quantum circuits as well as significantly reduces the node count needed in their implementation---especially for large systems.
The following example demonstrates the profound implications that this proposed change brings along.
\begin{example}
	Say that we want to recreate the situation in \autoref{ex:LargeBellState}, but completely strip away the identity nodes.
	Then, the operations become significantly more compact as illustrated in~\autoref{fig:BellState_NewDD}.
	Now, the single-qubit Hadamard gate is only represented by a single-level DD, while the two-qubit CNOT gate is represented by a two-level DD.
	Overall, this reduces the node count for these operations from $100$ and $199$ to $1$ and $2$, respectively.
\end{example}

\begin{algorithm}[t]
    \caption{Gate DD Creation (simplified)}
    \label{alg:makeGateDD}
    \begin{algorithmic}[1]
    \Procedure{GateDD}{$n$, $c$, $t$, $U$}
    \State $\begin{bsmallmatrix} e_{00} & e_{01}&e_{10} & e_{11}\end{bsmallmatrix} \gets$ \Call{Terminal}{}($\begin{bsmallmatrix} u_{00} & u_{01}&u_{10} & u_{11}\end{bsmallmatrix}$)
    \color{Red3}
    \For{$l$= 0,\dots,t-1} \Comment{Identities below $t$}
      \State $e_{ij} \gets$ \Call{Node}{}($l$, $\begin{bsmallmatrix} e_{ij} & 0 &0 & e_{ij}\end{bsmallmatrix}$)
    \EndFor
    \color{black}
	\State $e \gets$ \Call{Node}{}($t$, $\begin{bsmallmatrix} e_{00} & e_{01}& e_{10} & e_{11}\end{bsmallmatrix}$)\label{target}
    \color{Red3}
    \For{$l$= $t$+1,\dots,$c$-1}  \Comment{Identities between $t$ and $c$}
      \State $e \gets$ \Call{Node}{}($l$, $\begin{bsmallmatrix} e & 0 &0 & e\end{bsmallmatrix}$)
    \EndFor
    \color{black}
    \State $e \gets$ \Call{Node}{}($c$, $\begin{bsmallmatrix} \mathcolor{Red3}{\mathbb{I}_{c-1}}\mathcolor{Green3}{1} & 0 & 0 & e\end{bsmallmatrix}$) \label{control} 
     \color{Red3}
     \For{$l$= $c$+1,\dots,$n$-1}  \Comment{Identities above $c$}
      \State $e \gets$ \Call{Node}{}($l$, $\begin{bsmallmatrix} e & 0 &0 & e\end{bsmallmatrix}$)
    \EndFor
    \color{black}
    \Return $e$
    \EndProcedure
   \end{algorithmic}
\end{algorithm}

\begin{algorithm}[h]
    \caption{DD Node Creation (simplified)}
    \label{alg:makeDDNode}
    \begin{algorithmic}[1]
    \Procedure{Node}{$l, \begin{bsmallmatrix} e_{00} & e_{01}&e_{10} & e_{11}\end{bsmallmatrix}$}
    \State $e\coloneqq(e_{\text{node}}, e_{\text{weight}}) \gets ($\Call{GetNode}{}$(), 1)$
    \State $e_{\text{node}}.l \gets l$  
    \State $e_{\text{node}}.\text{edges} \gets \begin{bsmallmatrix} e_{00} & e_{01}&e_{10} & e_{11}\end{bsmallmatrix}$
    \State $e \gets$ \Call{Normalize}{$e$}
    \color{Green3}
    \If{\Call{ResemblesIdentity}{$e$}}
    \State $s \gets$ \Call{Successor}{$e$, $0$}
    \State \Call{Free}{$e_{\text{node}}$}
    \State \Return $s$
    \EndIf
    \color{black}
    \State \Return \Call{UtLookup}{$e$}
    \EndProcedure
   \end{algorithmic}
\end{algorithm}

\section{Implementation} \label{sec:Implementation}
While the change proposed above is seemingly small, the underlying assumption that two interacting DDs always have to act on the same number of levels is deeply rooted in all of the methods present in state-of-the-art realizations.
Hence, this section discusses the key changes required to deliver on the promise of stripping DDs of their identity---specifically in $(1)$ the creation of DDs for quantum operations, $(2)$ the creation of DD nodes themselves, and $(3)$ the DD operations such as multiplication and addition.
For each of these, a simplified pseudo-code diff is provided to illustrate the change.
To this end, additions will be marked in \textcolor{Green3}{green} while deletions will be marked in \textcolor{Red3}{red}.
The interested reader is welcome to check out the open source implementation at \url{https://github.com/cda-tum/mqt-core} for further details. 

The most obvious change lies in the method used to create DDs for quantum gates which, previously, had to be extended to the full system size by explicitly inserting identities for levels not acted on by the gate.
For simplicity, we only consider the case of a two-qubit controlled-$U$ gate ($U$ being specified as a $2\times 2$ unitary matrix) with control qubit $c$ and target qubit $t$ in an $n$-qubit system.
We additionally assume that the control qubit comes before the target qubit in the variable order of the DD, i.e., $t < c < n$. This is not required by the implementation, but is simpler to illustrate the algorithm.
Then, \autoref{alg:makeGateDD} sketches the corresponding method and how it was adapted to not even create the identities.
In the new implementation, the method only ever touches the levels the operation acts on---regardless of system size $n$.

However, it is not sufficient to simply avoid creating identity nodes during gate construction.
Such nodes may naturally occur as the result of a computation, e.g., in lines \ref{target} or \ref{control} in \autoref{alg:makeGateDD}.
Hence, it is also required to adapt the method for creating DD nodes given a level $l$ and a list of successor DDs $\begin{bsmallmatrix} e_{00} & e_{01}&e_{10} & e_{11}\end{bsmallmatrix}$.
This is shown in \autoref{alg:makeDDNode}.
Compared to the original implementation, an additional check is introduced after the normalization that triggers if the normalized DD node resembles the identity, i.e., its first and last successor point to the same node with an edge weight of one, while the other successors are zero.
If so, the newly created node is freed (as it is not needed) and the first successor is returned as a result of the call---effectively skipping the identity node.

Finally, the addition and multiplication routines were modified to account for the fact that two DDs that act as operands in these routines can no longer be assumed to always act on the same level due to potentially skipped nodes.
For illustrative purposes, we consider the multiplication algorithm here that takes a matrix DD $U$, a vector DD $v$, as well as a level $l$ and recursively computes the matrix-vector product $Uv$.
The resulting algorithm is shown in \autoref{alg:multiplication}.
Compared to the original version, it checks whether the matrix DD is at the correct level and implicitly treats the DD as an identity if it is not.
In the following, the implications of the above changes on the overall methodology are discussed---both in theoretical limits as well as practical performance.

\begin{algorithm}[t]
    \caption{Matrix-Vector Multiplication (simplified)}
    \label{alg:multiplication}
    \begin{algorithmic}[1]
        \Procedure{Multiply}{$U$, $v$, $l$}
        \If{\Call{IsZero}{$U$} or \Call{IsZero}{$v$}} \Return $0$ \EndIf
        \If{\Call{IsIdentity}{$U$}} \Return $v$ \EndIf
        \If{success, r $\gets$ \Call{CtLookup}{$U, v$}} \Return r \EndIf
        
        \State edges $\gets \begin{bsmallmatrix} 0 & 0 &0 & 0\end{bsmallmatrix}$
        
        \For{$i$, $j=0,1$}
        	\color{Green3}
        	\If{$U.l=l$}
        	\color{black}
      		\State $e_1 \gets$ \Call{Successor}{$U$, $2 i + j$}
			\color{Green3}
			\Else
			\; $e_1 \gets i=j\;?\; x\;\colon\; 0$ \EndIf
			\color{black}
      		\State $e_2 \gets$ \Call{Successor}{$y$, $j$}
      		\State $m \gets$ \Call{Multiply}{$e_1$, $e_2$, $l-1$}
      		\State $\text{edges}[i]\gets$\Call{Add}{$\text{edges}[i]$, $m$}
    	\EndFor
    	\State $r\gets$ \Call{Node}{$l$, edges}
    	\State \Call{CtInsert}{$U$, $v$, $r$}
    	\State \Return $r$
		\EndProcedure
   \end{algorithmic}
\end{algorithm}

\section{Implications} \label{sec:implications}


When developing data structures for quantum computing, the scalability with the number of qubits is one of the crucial criteria for determining a method's viability.
In the previous iterations of decision diagrams, all gates on an $n$-qubit system must be scaled to $n$-level DDs. 
This is objectively a significant bottleneck for scalability,
as infinite-level operations, i.e. ($n \rightarrow \infty$) would be required in the asymptotic limit. 
However, removal of the identity nodes means that the DDs representing gates are no longer connected to the
overall number of qubits, but rather the number of qubits the gate acts upon. 
This localization gives credence to the theoretical capabilities of DDs in representing very large systems.


Stripping identity nodes in DDs for operations also directly leads to a reduction in the overall node count and, consequently, the memory required to represent the necessary quantities for a particular task.
Instead of scaling with the overall system size, the memory requirements to store operations are entirely based on the number of qubits upon which they operate.
As confirmed by the experimental evaluations, which are summarized in \autoref{sec:Evaluation}, this leads to a drastic reduction in the overall number of allocated nodes.

The reduction in the number of nodes also has direct consequences on the performance of key data structures within the DD package.
This most significantly affects the \emph{unique table}, which is used to check whether two DD nodes represent the same functionality and, thus, to ensure canonicity of the data structure.
By reducing the number of nodes, there are significantly fewer lookups and inserts into this unique table (which is commonly implemented as a hash table for each variable). 
This implies that less cleanup (so-called \emph{garbage collection}) is required to get rid of superfluous entries and guarantee the amortized $O(1)$ complexity for lookup and insertion.
Since garbage collection is quite costly when it comes to runtime, this reduction in frequency constitutes a major performance improvement.
A similar impact applies to the compute table.



\begin{table*}[!t]
	\caption{Experimental Results}
	\label{tab:Evaluation}
    \centering
	\scalebox{0.7}{
    \begin{tabular}{@{} r @{\hspace{4em}} r r r r r r r r r r r r r r r r}
	 & \multicolumn{8}{c}{\textbf{Circuit Simulation}} & \multicolumn{8}{c}{\textbf{Unitary Simulation}} \\
	 \cmidrule(lr){2-9} \cmidrule(lr){10-17}
	 & & & \multicolumn{3}{c}{\textbf{Old}} & \multicolumn{3}{c}{\textbf{New}} & & & \multicolumn{3}{c}{\textbf{Old}} & \multicolumn{3}{c}{\textbf{New}} \\
	\cmidrule(lr){4-6} \cmidrule(lr){7-9} \cmidrule(lr){12-14} \cmidrule(lr){15-17}
	& \multicolumn{1}{r}{$n$} & \multicolumn{1}{r}{$|G|$} & \multicolumn{1}{r}{$t$ [s]} & \multicolumn{1}{r}{$|V|$} & \multicolumn{1}{r}{$|GC|$} & \multicolumn{1}{r}{$t$ [s]} & \multicolumn{1}{r}{$|V|$} & \multicolumn{1}{r}{$|GC|$} & \multicolumn{1}{r}{$n$} & \multicolumn{1}{r}{$|G|$} & \multicolumn{1}{r}{$t$ [s]} & \multicolumn{1}{r}{$|V|$} & \multicolumn{1}{r}{$|GC|$} & \multicolumn{1}{r}{$t$ [s]} & \multicolumn{1}{r}{$|V|$} & \multicolumn{1}{r}{$|GC|$} \\
	\midrule
	BV & 256 & 636 & <0.1 & 49 506 & 0 & <0.1 & 382 & 0 & 256 & 636 & 0.1 & 180 645 & 1 & <0.1 & 81 753 & 0 \\ 
        ~ & 512 & 1 271 & 0.3 & 328 054 & 2 & 0.1 & 761 & 0 & 512 & 1 271 & 0.4 & 788 888 & 6 & 0.2 & 327 378 & 2 \\ 
        ~ & 1 024 & 2 546 & 1.3 & 1 311 591 & 9 & 0.5 & 1 524 & 0 & 1 024 & 2 546 & 1.7 & 6 442 227 & 24 & 0.7 & 4 595 427 & 10 \\ 
        ~ & 2 048 & 5 104 & 6.9 & 5 252 986 & 39 & 2.7 & 3 058 & 0 & 2 048 & 5 104 & 8.7 & 25 854 080 & 97 & 3.7 & 7 420 378 & 40 \\ 
        ~ & 4 096 & 10 221 & 40.6 & 21 213 345 & 160 & 17.5 & 6 127 & 0 & 4 096 & 10 221 & 52.3 & 51 188 288 & 399 & 21.9 & 21 018 096 & 164 \\ 
        \vspace*{0.2cm} \\
		GHZ State & 256 & 256 & <0.1 & 33 151 & 0 & <0.1 & 511 & 0 & 256 & 256 & <0.1 & 125 792 & 0 & <0.1 & 33 151 & 0 \\ 
        ~ & 512 & 512 & 0.1 & 132 350 & 1 & <0.1 & 10 23 & 0 & 512 & 512 & 0.1 & 265 600 & 2 & 0.1 & 131 839 & 1 \\ 
        ~ & 1 024 & 1 024 & 0.5 & 529 915 & 4 & 0.2 & 2 047 & 0 & 1 024 & 1 024 & 0.5 & 1 049 600 & 8 & 0.3 & 525 823 & 4 \\ 
        ~ & 2 048 & 2 048 & 2.6 & 2 132 975 & 16 & 1.0 & 4 095 & 0 & 2 048 & 2 048 & 2.7 & 4 196 352 & 32 & 1.4 & 2 100 223 & 16 \\ 
        ~ & 4 096 & 4 096 & 15.6 & 8 660 926 & 65 & 6.6 & 8 191 & 0 & 4 096 & 4 096 & 16.3 & 17 875 872 & 131 & 8.0 & 8 394 751 & 64 \\ 
        \vspace*{0.2cm} \\
		W State & 256 & 1 021 & 0.1 & 132 093 & 1 & 0.1 & 1 786 & 0 & 256 & 1 021 & 0.3 & 651 639 & 4 & 0.2 & 520 971 & 3 \\ 
        ~ & 512 & 2 045 & 0.6 & 527 156 & 4 & 0.3 & 3 578 & 0 & 512 & 2 045 & 1.2 & 2 614 852 & 19 & 0.9 & 2 090 512 & 15 \\ 
        ~ & 1 024 & 4 093 & 2.7 & 2 106 893 & 16 & 1.3 & 7 162 & 0 & 1 024 & 4 093 & 5.8 & 10 477 875 & 80 & 4.5 & 8 375 330 & 64 \\ 
        ~ & 2 048 & 8 189 & 14.4 & 8 440 874 & 64 & 7.1 & 14 330 & 0 & 2 048 & 8 189 & 32.0 & 41 970 068 & 322 & 24.5 & 33 527 949 & 257 \\ 
        ~ & 4 096 & 16 381 & 85.6 & 33 923 956 & 256 & 46.2 & 28 666 & 0 & 4 096 & 16 381 & 200.1 & 168 183 059 & 1299 & 159.6 & 134 164 915 & 1 041 \\ 
        \vspace*{0.2cm} \\
		Grover & 28 & 1 019 143 & 0.1 & 45 146 & 0 & 0.1 & 42 668 & 0 & 28 & 1 019 143 & 0.2 & 9 580 & 0 & 0.2 & 7 345 & 0 \\ 
        ~ & 32 & 4 658 751 & 0.3 & 84 913 & 0 & 0.3 & 81 583 & 0 & 32 & 4 658 751 & 0.8 & 151 349 & 0 & 0.7 & 147 866 & 0 \\ 
        ~ & 36 & 20 964 167 & 1.2 & 156 491 & 1 & 1.1 & 152 108 & 1 & 36 & 20 964 167 & 2.6 & 266 061 & 1 & 2.5 & 260 593 & 1 \\ 
        ~ & 40 & 93 174 159 & 1.1 & 272 885 & 3 & 0.9 & 268 040 & 3 & 40 & 93 174 159 & 3.1 & 429 992 & 3 & 2.6 & 427 986 & 3 \\ 
        ~ & 42 & 195 665 483 & 1.5 & 238 364 & 1 & 1.3 & 233 159 & 1 & 42 & 195 665 483 & 1.6 & 336 721 & 2 & 1.4 & 328 534 & 2 \\ 
        \vspace*{0.2cm} \\
		QFT & 256 & 33 024 & 1.0 & 1 435 826 & 10 & 0.1 & 20 380 & 0 & 18 & 180 & 8.4 & 527 414 & 4 & 0.3 & 524 591 & 4 \\ 
        ~ & 512 & 131 584 & 7.5 & 5 785 838 & 44 & 0.7 & 42 652 & 0 & 19 & 199 & 49.1 & 1 052 288 & 7 & 1.1 & 1 048 915 & 7 \\ 
        ~ & 1 024 & 525 312 & 67.1 & 23 251 301 & 176 & 3.5 & 87 196 & 0 & 20 & 220 & 338.8 & 2 101 530 & 11 & 6.4 & 2 097 529 & 11 \\ 
        ~ & 2 048 & 2 099 200 & 566.9 & 93 733 455 & 711 & 21.3 & 176 284 & 1 & 21 & 241 & 2 456.5 & 4 199 433 & 16 & 38.6 & 4 194 721 & 16 \\ 
        ~ & 4 096 & 8 392 704 & 4 683.8 & 380 202 897 & 2 870 & 134.5 & 354 460 & 2 & 22 & 264 & 18 855.5 & 8 394 578 & 22 & 267.8 & 8 389 067 & 22 \\ 
        \vspace*{0.2cm} \\
		QPE & 15 & 134 & 0.2 & 1 254 & 0 & 0.2 & 225 & 0 & 8 & 43 & <0.1 & 10 676 & 0 & <0.1 & 10 507 & 0 \\ 
        ~ & 16 & 151 & 1.7 & 1 527 & 2 & 1.2 & 257 & 2 & 9 & 53 & <0.1 & 41 760 & 1 & <0.1 & 41 519 & 1 \\ 
        ~ & 17 & 169 & 27.3 & 1 835 & 3 & 9.1 & 291 & 3 & 10 & 64 & 0.3 & 165 727 & 2 & 0.2 & 165 387 & 2 \\ 
        ~ & 18 & 188 & 333.8 & 2 180 & 4 & 57.0 & 327 & 4 & 11 & 76 & 13.4 & 661 130 & 5 & 3.7 & 660 649 & 5 \\ 
        ~ & 19 & 208 & 2 239.5 & 2 564 & 5 & 332.8 & 365 & 5 & 12 & 89 & 667.8 & 2 643 492 & 8 & 82.9 & 2 641 229 & 8  \\
    \bottomrule
	\end{tabular}
	}
	\vspace*{1mm}
	{\\ \small $n$: Number of qubits \hspace*{0.2cm} $|G|$: Gate count \hspace*{0.2cm} $t$ [s]: Runtime \hspace*{0.2cm} $|V|$: Node Count \hspace*{0.2cm} $|GC|$: Garbage Collection Runs}
\end{table*}

\section{Experimental Evaluations} \label{sec:Evaluation}

In order to evaluate the impact of the newly proposed type of decision diagrams, we implemented the removal of identity nodes on top of the state-of-the-art decision diagram package publicly available as part of the MQT Core library (\url{https://github.com/cda-tum/mqt-core}). The resulting simulator is available as part of the \emph{Munich Quantum Toolkit} (MQT; \cite{wille_mqt_2024}) at \url{https://github.com/cda-tum/mqt-ddsim}.
Then, we benchmarked the resulting implementation against the original package by performing statevector simulations (in which the gates are applied sequentially to the initial state) and unitary simulations (in which the functionality of the quantum algorithm is constructed directly) for a wide range of benchmarks---including generating the GHZ state, generating the W state,
the Bernstein-Vazirani (BV) algorithm, Quantum Fourier Transform (QFT), Quantum Phase Estimation (QPE), and Grover's algorithm.
For both types of simulation, each algorithm is simulated for several different qubit counts $n$, while the gate count $|G|$, runtime $t$, overall count of matrix DD nodes $|V|$, as well as number of garbage collection runs $|GC|$ of the old and new implementation are recorded.
These results are summarized in~\autoref{tab:Evaluation}.

The results align well with the theoretical and practical expectations laid out in~\autoref{sec:implications}.
As expected, removing the identity nodes yields an increase in performance across all benchmarks---resulting in an average speed-up of $7.7\times$ and up to a $70\times$ improvement compared to the state of the art.
Furthermore, the overall node count required for the simulation is reduced, on average, by a factor of $218\times$ and up to $3462\times$.
This reduction in the number of nodes allows the new implementation to perform fewer garbage collections, which, along with the reduction in node count and required operations, is a major factor for the drastic runtime decrease that can be observed for both types of simulation.

\section{Conclusion} \label{sec:Conclusion}
By stripping quantum decision diagrams of their identity nodes, decision diagrams have been brought closer to a natural, more efficient representation of quantum operations.
In this work, it has been shown that this change is not only theoretically motivated towards allowing DDs to push towards larger and larger systems, but also sees significant practical advantage
compared to previous state-of-the-art implemenations. Due to the significant runtime and storage benefits presented here, it is expected that the structure of the decision diagrams outlined in this work will supplement previous implementations to become
the de facto representation used to simulate, verify, and compile quantum circuits.

\section*{Acknowledgments}
This work received funding from the European Research Council (ERC) under the European Union’s Horizon 2020 research and innovation program (grant agreement No. 101001318), was part of the Munich Quantum Valley, which is supported by the Bavarian state government with funds from the Hightech Agenda Bayern Plus, and has been supported by the BMWK on the basis of a decision by the German Bundestag through project QuaST.

\printbibliography

\end{document}